# Transformation-optics modeling of 3D-printed freeform waveguides

Aleksandar Nesic,[1,*] Matthias Blaicher,[1,2] Emilio Orlandini,[1] Tudor Olariu,[1] Maria Paszkiewicz,[3] Fernando Negredo,[3] Pascal Kraft,[4] Mariia Sukhova,[4] Andreas Hofmann,[5] Willy Dörfler,[4] Carsten Rockstuhl,[3,6] Wolfgang Freude,[1] and Christian Koos[1,2,**]

[1] *Karlsruhe Institute of Technology (KIT), Institute of Photonics and Quantum Electronics (IPQ), Engesserstrasse 5, 76131 Karlsruhe, Germany*
[2] *Karlsruhe Institute of Technology (KIT), Institute of Microstructure Technology (IMT), Hermann-von-Helmholtz-Platz 1, 76344 Eggenstein-Leopoldshafen, Germany*
[3] *Karlsruhe Institute of Technology (KIT), Institute of Theoretical Solid State Physics (TFP), Wolfgang-Gaede-Strasse 1, 76131 Karlsruhe Germany*
[4] *Karlsruhe Institute of Technology (KIT), Institute for Applied and Numerical Mathematics (IANM), Englerstrasse 2, 76131 Karlsruhe Germany*
[5] *Karlsruhe Institute of Technology (KIT), Institute for Automation and Applied Informatics (IAI), Hermann-von-Helmholtz-Platz 1, 76344 Eggenstein-Leopoldshafen, Germany*
[6] *Karlsruhe Institute of Technology (KIT), Institute of Nanotechnology (INT), Hermann-von-Helmholtz-Platz 1, 76344 Eggenstein-Leopoldshafen, Germany*
[*] *aleksandar.nesic@kit.edu*, [**] *christian.koos@kit.edu*

**Abstract:** Multi-photon lithography allows to complement planar photonic integrated circuits (PIC) by in-situ 3D-printed freeform waveguide structures. However, design and optimization of such freeform waveguides using time-domain Maxwell's equations solvers often requires comparatively large computational volumes, within which the structure of interest only occupies a small fraction, thus leading to poor computational efficiency. In this paper, we present a solver-independent transformation-optics-(TO-) based technique that allows to greatly reduce the computational effort related to modeling of 3D freeform waveguides. The concept relies on transforming freeform waveguides with curved trajectories into equivalent waveguide structures with modified material properties but geometrically straight trajectories, that can be efficiently fit into rather small cuboid-shaped computational volumes. We demonstrate the viability of the technique and benchmark its performance using a series of different freeform waveguides, achieving a reduction of the simulation time by a factor of 3–6 with a significant potential for further improvement. We also fabricate and experimentally test the simulated waveguides by 3D-printing on a silicon photonic chip, and we find good agreement between the simulated and the measured transmission at $\lambda = 1550$ nm.

## 1. Introduction

Photonic integration has evolved into a key technology for a wide variety of applications that range from high-speed communications [1], ultra-fast signal processing [2, 3] and artificial intelligence [4] to optical metrology and sensing [5–8] and to biophotonics and life sciences [9–11]. On the technological level, photonic integrated circuits (PIC) predominantly rely on planar structures that can be fabricated with well-established microfabrication techniques based on layer deposition and 2D patterning via high-resolution electron-beam or deep-UV lithography. More recently, these techniques have been complemented by multi-photon lithography that allows for in-situ fabrication of functional 3D freeform structures that can greatly enhance the functionality and versatility of planar PIC. Examples are 3D-printed chip-chip connections, so-called photonic wire bonds [12, 13] that open an attractive path towards high-performance hybrid multi-chip modules [14], 3D-printed waveguide overpasses [15], reconfigurable photonic circuits [16, 17], 3D-printed waveguide interconnects for photonic neural networks [18], 3D-printed waveguide splitters [19], or 3D-printed polarization splitters and rotators [20]. However, while simulation tools for planar lightwave structures are available as part of commercial software packages [21–24], efficient modeling and design of 3D freeform waveguides still represents a challenge. This applies in particular to numerical solvers that rely on rectilinear grids within cuboid-shaped computational domains, which is, e.g., the case for most time-domain techniques. Applying such solvers to 3D freeform waveguides with strongly curved non-plane trajectories requires comparatively large computational volumes, within which the structure of interest only occupies a small fraction, thus leading to poor computational efficiency. In addition, accurate representation of curved surfaces on rectilinear grids requires local refinement of the mesh cells, which leads to reduced time steps in time-domain simulations and thus increases overall simulation time.

Here, we present a transformation-optics (TO) method for reducing the computational effort associated with freeform-waveguide simulations. Our approach relies on transforming curved freeform waveguides in the original 3D space into straight waveguides in a virtual 3D space, which can then be efficiently treated by rigorous time-domain Maxwell's equations solvers defined on a rectilinear grid. Specifically, the transformed waveguide in the virtual space can be confined in a rectangular simulation box, whose volume is comparable to the actual freeform waveguide volume. Furthermore, in case of freeform waveguides with rectangular cores, the grid lines in the virtual space are perfectly aligned with the core surfaces thereby eliminating the need for any local mesh cell refinement. We demonstrate the viability of the concept using the commercially available time-domain solver of CST Microwave Studio® (CST MWS), which is based on the finite integration technique (FIT) [25, 26], reaching an acceleration by a factor of 3–6. In addition, we fabricate the simulated freeform waveguides on a silicon photonic (SiP) chip and measure the transmission losses at a vacuum wavelength of $\lambda = 1550$ nm. We find excellent agreement between TO-based simulations in virtual space and the associated reference simulations in real space, and we confirm that the simulated transmission losses match their experimentally measured counterparts reasonably well. Although primarily aimed for use with time-domain solvers on a rectilinear grid, our method represents a general technique that allows to transform freeform waveguide-based devices into straight structures and is independent of the underlying solver.

## 2. TO based concept of freeform waveguide modeling

The TO concept relies on the fact that Maxwell's equations are form-invariant with respect to coordinate transformations. In particular, if we map an *original* domain from an $(x, y, z)$-coordinate system, to a *virtual* domain in a $(u, v, s)$-coordinate system, we only need to adapt the material properties in the virtual domain, while the form of Maxwell's equations remains unchanged [27–30]. In case of a coordinate transformation described by a differentiable function $(u, v, s)^T = \mathbf{f}(x, y, z)$, where $(x, y, z), (u, v, s) \in \mathbb{R}^3$, the relationship between the material properties in the original $(x, y, z)$-space and in the virtual $(u, v, s)$-space reads [27–30]:

$$\varepsilon'(u, v, s) = \frac{\mathbf{J}(x, y, z) \cdot \varepsilon(x, y, z) \cdot \mathbf{J}^T(x, y, z)}{\det(\mathbf{J}(x, y, z))},$$
$$\mu'(u, v, s) = \frac{\mathbf{J}(x, y, z) \cdot \mu(x, y, z) \cdot \mathbf{J}^T(x, y, z)}{\det(\mathbf{J}(x, y, z))}. \tag{1}$$

In these relations, the quantities $\varepsilon$ and $\varepsilon'$ denote the dielectric permittivity tensors, $\mu$ and $\mu'$ are the magnetic permeability tensors, and $\mathbf{J}$ is the Jacobian matrix of the coordinate transformation function **f**,

$$\mathbf{J} = \begin{bmatrix} \frac{\partial u}{\partial x} & \frac{\partial u}{\partial y} & \frac{\partial u}{\partial z} \\ \frac{\partial v}{\partial x} & \frac{\partial v}{\partial y} & \frac{\partial v}{\partial z} \\ \frac{\partial s}{\partial x} & \frac{\partial s}{\partial y} & \frac{\partial s}{\partial z} \end{bmatrix}. \quad (2)$$

Note that the transformed media in virtual space are generally anisotropic and magnetic, even if the structure in original space is made from isotropic non-magnetic material.

The TO concept has previously been used to analyse and design a variety of devices such as for beam deflectors and expanders [31], polarization splitters and rotators [32], flat lenses [33], multimode waveguide bends [34], or electromagnetic cloaks for hiding of objects [35–37] or for reshaping the perception of cloaked objects [38, 39]. In general it is possible to design TO-based devices with arbitrary shapes [40], and the TO concept can be further expanded into plasmonics [41], thermodynamics and mechanics [42], and even beyond, including spacetime and general relativity [43]. Here, we exploit TO for efficient numerical modeling of 3D freeform waveguides using well-established time-domain solvers, see Fig. 1 for an illustration of the underlying concept. Generally, time-domain solvers relying, e.g., on finite-difference-time-domain (FDTD) techniques, are perfectly suited for large-scale simulations, offering a numerical complexity that scales linearly with problem size while being amenable to efficient parallelization on large-scale computer clusters. Moreover, time-domain techniques are robust, lend themselves to broadband or transient simulations, and even offer a natural path to handling of nonlinear behavior [44]. On the other hand, time-domain techniques usually rely on rather inflexible rectilinear grids and cuboid-shaped computational domains, which severely limits the performance when applied to 3D freeform waveguides. Specifically, the rectilinear grid does not allow for efficient representation of curved surfaces, while the cuboid-shaped computational domain leads to poor computational efficiency with comparatively large computational volumes within which the structure of interest only occupies a small fraction, see Fig. 1(a).

To overcome these problems, we map the freeform waveguide with a curved trajectory in the original $(x, y, z)$-space to an equivalent waveguide with a straight trajectory and modified permittivity and permeability tensors in a virtual $(u, v, s)$-space, see Fig. 1(b). In the virtual space, we may then use a rectangular computational domain that only encompasses the straight waveguide and its direct vicinity, along with a rectilinear grid that is oriented along the direction of the waveguide. The virtual waveguide can thus be efficiently modeled by a conventional time-domain solver, and the results are then transformed back to the $(x, y, z)$-coordinate system to obtain the field distributions in original space.

To implement this technique, we need to define a function $(u, v, s)^{\mathrm{T}} = \mathbf{f}(x, y, z)$ that transforms the curved waveguide in original space into a straight path in virtual space. It is actually easier and more intuitive to analytically express the inverse function $(x, y, z)^{\mathrm{T}} = \mathbf{f}^{-1}(u, v, s)$ that maps a point $(u, v, s)$ in transformed space back to original space. To arrive at a mathematical formulation of $\mathbf{f}^{-1}$, we assign the coordinate $s$ to the arc length of the waveguide trajectory $\mathbf{r}(s) = (x_0(s), y_0(s), z_0(s))$ in original space, while $u$ and $v$ are associated with the transverse position relative to the waveguide trajectory, where the direction is defined by a pair of unit vectors $\mathbf{U}$ and $\mathbf{V}$. This leads to the relation

$$\begin{bmatrix} x(u, v, s) \\ y(u, v, s) \\ z(u, v, s) \end{bmatrix} = \mathbf{f}^{-1}(u, v, s) = \begin{bmatrix} x_0(s) \\ y_0(s) \\ z_0(s) \end{bmatrix} + u\mathbf{U}(s) + v\mathbf{V}(s). \quad (3)$$

The $s$-coordinate, i.e., the arc length of the waveguide trajectory is defined such that $s = 0$ in its starting point, and the unit vectors $\mathbf{U}$ and $\mathbf{V}$ are chosen such that they form a right-handed trihedron $(\mathbf{U}, \mathbf{V}, \mathbf{T})$ with the tangent vector $\mathbf{T} = \dfrac{\mathrm{d}\mathbf{r}}{\mathrm{d}s}$. Note that the trajectory $\mathbf{r}$ is generally not parametrized with respect to its arc length $s$, but with respect to some other parameter $t$, with $\mathbf{r}(t) = (x(t), y(t), z(t))$. The two parametrizations are connected by the relation $s = \int_0^\tau \left|\frac{\mathrm{d}\mathbf{r}}{\mathrm{d}t}\right| \mathrm{d}t$.

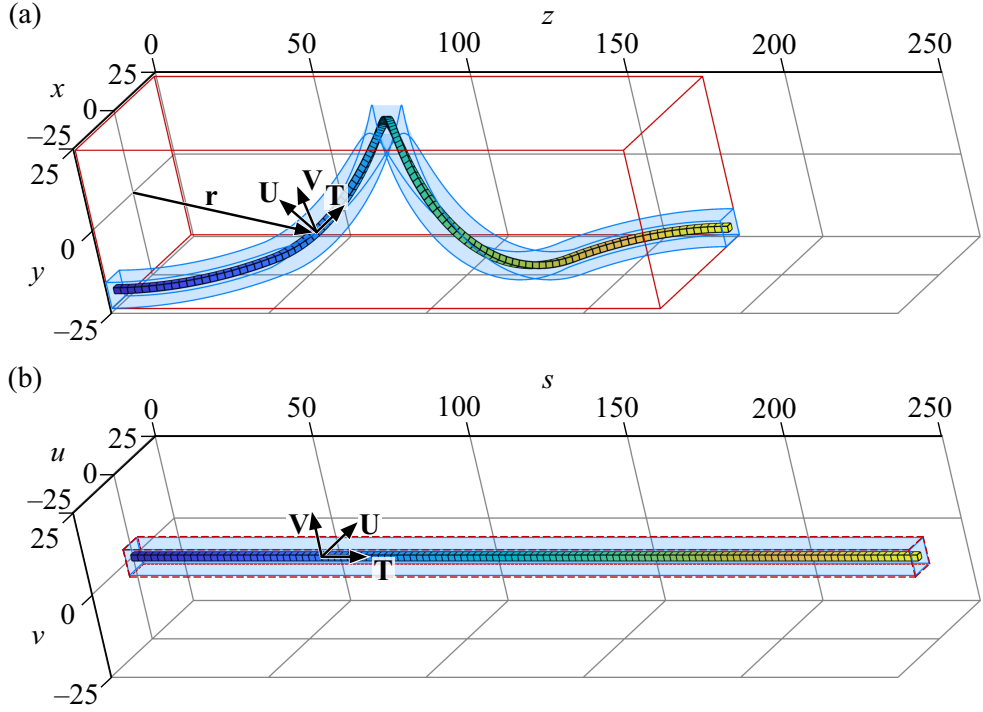

Fig. 1. Transformation of a freeform optical waveguide from the original $(x, y, z)$-space to the virtual $(u, v, s)$-space. **(a)** Sample freeform waveguide in the original $(x, y, z)$-space. The computational domain necessary to simulate the structure with a time-domain solver on a rectilinear grid is determined by the rectangular red box. However, the region of interest, i.e., the relevant part of the computational domain, only comprises the close vicinity of the freeform waveguide — this part and its edges are depicted in blue. The rest of the computational domain is unimportant and unnecessarily consumes computing resources. Moreover, a fine discretization would be needed to correctly represent the curved surfaces of the freeform waveguide by a rectilinear computational grid. The coordinate transformation described in Eq. (3) is defined by the unit vectors $\mathbf{U}$, $\mathbf{V}$, and $\mathbf{T}$, where $\mathbf{T}$ is the local tangent unit vector $\mathrm{d}\mathbf{r}/\mathrm{d}s$ of the trajectory, while $\mathbf{U}$ and $\mathbf{V}$ span the transverse plane in the respective trajectory point. The vectors are chosen such that $(\mathbf{U}, \mathbf{V}, \mathbf{T})$ is a right-handed trihedron. **(b)** Transformed waveguide in the virtual $(u, v, s)$-space. The freeform waveguide trajectory in the original space has been mapped to a straight line, and the relevant part of the computational domain that was "twisted" in the original $(x, y, z)$-space is now represented by a rectangular blue box. The computational domain can now be reduced to the region of interest, see red dashed rectangular box. The coordinate transformation significantly reduces the computational domain at the cost of pre-calculating the spatial distributions of tensors of dielectric permittivity and magnetic permeability in $(u, v, s)$-space. Moreover, the rectilinear computational grid may be better adapted to the shape of the waveguide, in particular for rectangular cross sections.

Equation (3) still leaves the freedom of choosing the orientation of the unit vectors $\mathbf{U}$ and $\mathbf{V}$ within the transverse plane in the respective trajectory point. One obvious choice for the $(\mathbf{U}, \mathbf{V}, \mathbf{T})$ frame could be the natural Frenet-Serret frame, where $\mathbf{U}$ could be chosen as the binormal vector, and **V** could be chosen as the normal vector of the trajectory **r** $(t)$. However, the Frenet-Serret frame is not the best choice because the binormal and the normal vectors are neither defined in points where the trajectory is straight, nor in inflection points. Furthermore, the frames immediately before and immediately after an inflection point are rotated against each other by 180° about the tangent vector. We therefore use the *rotation-minimizing frame* (RMF) [45, 46], which minimizes the frame spinning along the trajectory, and which is commonly used in computer graphics and 3D modeling [47]. To calculate the RMF, we use a simple and fast approximation method called *double reflection method* [48]. This method requires the trajectory sample points, the tangent vectors **T** in all sample points, and a coordinate frame in the first sample point $(\mathbf{U}_1, \mathbf{V}_1, \mathbf{T}_1)$ as an input. The coordinate frames in the remaining sample points along the trajectory are calculated recursively [48]. The coordinate frame $(\mathbf{U}_1, \mathbf{V}_1, \mathbf{T}_1)$ in the first sample point is chosen such that $\mathbf{T}_1$ is the tangent, and the two remaining mutually perpendicular vectors $\mathbf{U}_1$ and $\mathbf{V}_1$ can be chosen arbitrarily in the plane that is perpendicular to $\mathbf{T}_1$.

Note that the coordinate transformation described in Eq. (3) is generally not bijective. A trivial example is the case of self-intersecting trajectories, which must be excluded. Another obvious problem might arise if two waveguide segments pass by each other or overpass each other within a close distance, e.g., in spirals, helices, and loops. In this case, care must be taken to avoid mapping the same sub-domain of the $(x, y, z)$-space being in the vicinity of both waveguide segments twice into two distinct sub-domains of the $(u, v, s)$-space. Finally, bijectivity might be violated if the local curvature of the trajectory is too strong, such that the center of curvature falls into the transformed domain. Specifically, for a given trajectory point, the ranges for $u$ and $v$, $u \in [u_{min}, u_{max}]$ and $v \in [v_{min}, v_{max}]$ should be chosen not to include the center of curvature for this specific trajectory point. In other words: If, in a certain trajectory point, the center of the curvature would be represented by $(u_c, v_c)$ in transformed space, then the range of $u$ and $v$ for this trajectory point must be limited not to include this point, which is ensured by the sufficient condition $(u_c, v_c) \notin \{(u, v) \mid u_{min} \leq u \leq u_{max} \,\wedge\, v_{min} \leq v \leq v_{max}\}$. For more details, see Appendix A.

## 3. Implementation of TO modeling of freeform waveguides

To prove the viability of the proposed approach, we implement and use the TO concept to simulate freeform waveguides with an invariant rectangular cross section of the core. For the coordinate transform, we use MATLAB®, and the resulting waveguide in virtual space is then treated with CST MWS, which relies on a time-domain finite-integration technique (FIT) [25, 26]. We restricted the implementation to freeform waveguides that are made from isotropic material and that feature plane trajectories, such that the tensors $\boldsymbol{\varepsilon}'$ and $\boldsymbol{\mu}'$ are diagonal in the transformed space, see Appendix B. Anisotropic materials and/or non-plane trajectory would have lead to non-diagonal tensors $\boldsymbol{\varepsilon}'$ and $\boldsymbol{\mu}'$, which cannot be treated by the FIT solver of CST MWS (CST Studio Suite version 2019).

The input data for our MATLAB® code consist of the trajectory sample points $(x_0, y_0, z_0)$ in the original $(x, y, z)$-space, the orientation of the RMF in the initial point, the cross-sectional shape of the waveguide core along with the material properties of the core and the cladding ($\varepsilon_{\text{core}}$, $\mu_{\text{core}}$, $\varepsilon_{\text{cladding}}$, and $\mu_{\text{cladding}}$), and the ranges of $u$- and $v$-coordinates that define the space to be transformed. Without loss of generality, we can assume that the plane trajectory lies in the $(y, z)$-plane, i.e., that the $x$-coordinates of all trajectory points are equal to zero. For simplicity, we chose the right-handed RMF $(\mathbf{U}_1, \mathbf{V}_1, \mathbf{T}_1)$ in the first sample point of the trajectory such that $\mathbf{U}_1$ is parallel to the $x$-axis and $\mathbf{V}_1$ is perpendicular to $\mathbf{U}_1$ and to the tangent vector $\mathbf{T}_1$. We further assume a rectangular core cross section with width $w_{\text{WG}}$ and height $h_{\text{WG}}$, measured along the **U**- and **V**-directions of the RMF. In a first step of the transformation, we numerically calculate the $s$-coordinate sample points from the trajectory sample points. The RMF $(\mathbf{U}(s), \mathbf{V}(s), \mathbf{T}(s))$ in the remaining trajectory points is then calculated numerically by the aforementioned double reflection method [48]. In the special case of a plane trajectory in the $(y, z)$-plane with $\mathbf{U}_1$ chosen to be parallel to the $x$-axis, $\mathbf{U}(s)$ is parallel to the $x$-axis in all trajectory points. Having calculated the $s$-coordinates with given $u$- and $v$-coordinate ranges, we have the rectangular box defining the computational domain in $(u, v, s)$-space. The ranges of $u$ and $v$ determine how much space around the trajectory will be taken into account for the TO-based simulations. These ranges should be large enough to include a sufficient portion of the evanescent field outside the core, but not too large, to ensure bijectivity of the function $(u, v, s)^{\mathrm{T}} = \mathbf{f}(x, y, z)$, see Appendix A.

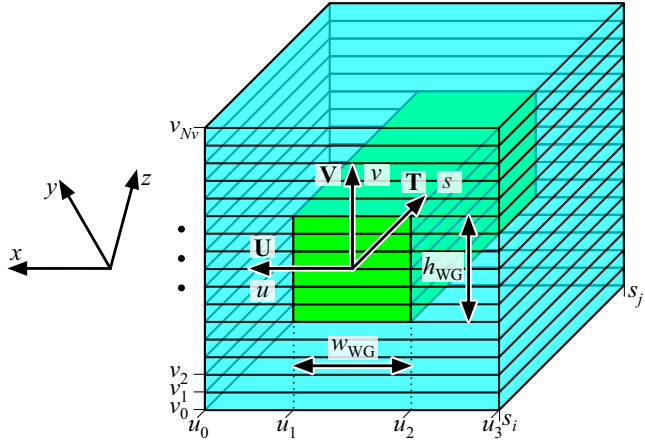

Fig. 2. Representation of a waveguide section with a trajectory lying in the $(y, z)$-plane by rectangular *bricks* with constant material properties in virtual $(u, v, s)$-space. The freeform waveguide is discretized into sections with approximately constant curvature of the waveguide trajectory (*slices*) along $s$, defined by coordinates $s_i$ and $s_j$. Green: Core region of the transformed freeform waveguide. Cyan: Cladding region of the transformed freeform waveguide. For a freeform waveguide with a plane trajectory in the $(y, z)$-plane, the axis **U** of the RMF is parallel to the $x$-axis in all trajectory sample points, which allows us to completely omit the discretization along $u$ above and below the freeform waveguide core and to reduce the representation along $u$ to three segments in the region of the waveguide core. The illustrated freeform waveguide slice is represented by an overall of 28 bricks — 5 each above and below the core, 6 on the each right and the left of the core, and 6 within the core.

In the next step, the transformed material properties in virtual space need to be calculated and fed to the numerical solver. Generally, this can be done by adapting $\varepsilon'$ and $\mu'$ in each point of space according to Eq. (1). For commercially available simulation programs with CAD-type user interfaces, however, setting the spatial variations of the materials properties on the level of individual spatial grid cells is not very efficient. We therefore approximate the transformed structure with continuously varying material properties by a multitude of small *bricks* with constant material properties, which can be fed to the solver as cuboid CAD elements by a standard scripting interface. We partition the range of $u$-coordinates into $N_u$ steps, the range of $v$-coordinates into $N_v$ steps, and the range of $s$-coordinates into $N_s$ steps. Then the total number of bricks is given by

$$N_{\mathrm{bricks}} = N_u N_v N_s. \tag{4}$$

Regarding the choice of the brick size, there is obviously a trade-off — too fine a discretization will require more time to calculate the material properties and more memory to store them, while too coarse a discretization might cause inaccurate simulation results. For better orientation, we provide a few rules of thumb that might help to select appropriate brick sizes. Note that these rules of thumb cannot replace a systematic convergence study for the respective waveguide cross section and for the curvature range of interest. In the following, we refer to the discretization along $s$ as *slicing*. Slicing the freeform waveguide in each sample point of the trajectory would result in a large number of *slices* $N_s$. We can reduce this number by exploiting the fact that each slice represents a section of the original waveguide having a constant bend radius. In a first step, we therefore calculate the bend radii in each trajectory sample point $R(s)$, which is done numerically. We then step through the trajectory points and calculate the change $\Delta R$ of the radius of curvature with respect to the first trajectory point. When the relative change $\Delta R/R$ exceeds a given threshold of, e.g., 0.1, we merge all preceding points into the first slice. We then repeat the procedure by using the last sample point of the first slice as a new reference for calculating the relative change $\Delta R/R$. The last slice is terminated by the last trajectory point of the waveguide. For proper choice of the brick sizes $\delta u$ and $\delta v$ along the transverse $u$ and $v$ direction, we need to make sure that the associated discretization of the dielectric and magnetic material properties does not introduce excessive perturbations of the optical wavefronts. As a rule of thumb we may require that, within the core and within the cladding, the difference $\delta \varepsilon'_{ij}$ and $\delta \mu'_{ij}$ of any two corresponding transformed $\varepsilon'$- and $\mu'$-tensor elements in any two neighboring bricks along $u$ and $v$ should be small compared to the difference of $\Delta \varepsilon$ between the core and the cladding of the original waveguide,

$$\left( \delta \varepsilon'_{ij} \ll \Delta \varepsilon \quad \wedge \quad \delta \mu'_{ij} \ll \Delta \varepsilon \right) \qquad \forall \, i, j \in \{1, 2, 3\} . \tag{5}$$

Note that, for waveguides with strong variations of the curvature along the trajectory, it might be difficult to fulfill the inequality according to Eq. (5) in all grid points that are contained in a rectangular bounding box in transformed space. It should then be ensured that Eq. (5) applies at least to the regions that bear significant electric fields. A systematic convergence study for the respective case of interest might be unavoidable to ensure proper representation of the waveguide structure.

Assuming a simplified waveguide structure with a plane trajectory that entirely lies in the $(y,z)$-plane allows to greatly simplify the Jacobian according to Eq. (2), see Eq. (9) of Appendix B. In this case, regions for which the material properties of the original waveguide in original space do not change along $x$ do not need to be subdivided into bricks along $u$. Since we consider a waveguide with homogeneous core and cladding region, we may thus reduce the transformed structure to $N_u^{\text{core}} = 3$ bricks along $u$ in the core region and $N_u^{\text{cladding}} = 1$ brick along $u$ in the cladding region below and above the core, see Fig. 2. The range of $v$-coordinates is divided into $N_v = N_v^{\text{core}} + N_v^{\text{cladding}}$ steps, which comprises $N_v^{\text{core}}$ steps in the core region, and $N_v^{\text{cladding}}$ steps in the cladding region below and above the core. Therefore, the number of bricks per slice in this case is $N_u^{\text{core}} N_v^{\text{core}} + N_u^{\text{cladding}} N_v^{\text{cladding}}$, and Eq. (4) becomes

$$N_{\text{bricks}}^{\text{plane trajectory}} = \left( N_u^{\text{core}} N_v^{\text{core}} + N_u^{\text{cladding}} N_v^{\text{cladding}} \right) N_s. \tag{6}$$

The example shown in Fig. 2 corresponds to the discretization that we did to perform simulations in the $(u, v, s)$-space, featuring $N_u^{\text{core}} = 3$, $N_u^{\text{cladding}} = 1$, $N_v^{\text{core}} = 6$, and $N_v^{\text{cladding}} = 5 + 5 = 10$ steps. This corresponds to $3 \cdot 6 + 1 \cdot 10 = 28$ bricks per slice. The total number of bricks is therefore $28 N_s$.

Finally, to calculate the material properties of the various bricks in $(u, v, s)$-space, we first need to know the material properties of the corresponding bricks in $(x, y, z)$-space and then transform them using Eq. (1). After calculating the material properties of all bricks, the MATLAB® code then writes a script which is loaded and run in CST MWS to generate the transformed structure for simulation by the time-domain solver of CST MWS. After the simulation is done, scattering parameters (S-parameters) of the freeform waveguide can be read directly, while the field distribution must additionally undergo the inverse coordinate transformation given by Eq. (3), in order to be represented in the original $(x, y, z)$-space. This is done by exporting the field distribution from the CST MWS simulation, and performing the inverse transformation by an additional MATLAB® script. Note that our technique leaves the underlying solver unchanged and may also allow to exploit existing FDTD implementations [21–24] to efficiently treat 3D freeform waveguides that comprise nonlinear or dispersive media [49, 50].

## 4. Freeform waveguide simulations and experimental benchmarking

To verify our TO approach, we apply it to a series of freeform waveguide structures that connect two SiP waveguides, see Fig. 3(b). We calculate the transmission and the electric field distribution for these structures at a wavelength of $\lambda = 1550$ nm using the TO concept with CST MWS as a FIT time-domain solver for the waveguide in the virtual $(u, v, s)$-space, and we compare the results to conventional simulations of the original structure in $(x, y, z)$-space. Moreover, we experimentally realized and characterized the structures — Figure 3 shows the experimental setup, some freeform waveguide illustrations, scanning electron microscopy (SEM) pictures of the fabricated structures, as well as a comparison of the simulated and measured transmission for different waveguide trajectories, that are described by the height $h$ of the apex above the substrate. As a further reference, we also calculated the transmission using our previously reported fundamental-mode approximation (FMA) method [51]. The FMA is based on a look-up table of pre-calculated eigenmodes (propagation constants and modal field distributions) of waveguide segments with constant radii of curvature, on their tabulated bending losses and on transition losses due to modal field mismatch between two adjacent segments with different radii of curvature. The FMA subdivides freeform waveguides in segments with constant radii of curvature and calculates the transmission losses based on the look-up table. While exceptionally fast, this method disregards the existence of higher-order modes.

For the experimental benchmark, the freeform waveguides were 3D-printed on a SiP chip, Figs. 3(b)–3(e), that was fabricated through a commercial foundry using standard CMOS process and 248 nm deep-UV lithography. The waveguides are normally covered by a $SiO_2$ top cladding layer, which was locally removed to make the SiP waveguides accessible to the 3D printing system. For fabrication of the 3D freeform waveguides, a negative-tone photoresist is deposited on the chip and structures are then 3D-printed by two-photon polymerization [52]. Subsequently, the unexposed resist is removed in a separate development step, and the freeform waveguides are covered by a low-index polymer (not shown in Fig. 3) that serves as a cladding and a protection against environmental influences. The refractive index of the freeform waveguide core at $\lambda = 1550$ nm amounts to $n_{core} \approx 1.53$, and the cladding refractive index is $n_{cladding} \approx 1.36$. Both materials are assumed to be lossless dielectrics (relative magnetic permeability $\mu_r = 1$ in the original $(x, y, z)$-space), and the corresponding values of the dielectric permittivity are calculated by squaring the refractive indices. Each freeform waveguide core has a $w_{WG} \times h_{WG} = 2\,\mu m \times 1.8\,\mu m$ rectangular cross-section and bridges a $l_g = 100\,\mu m$ gap between a pair of SiP strip waveguides (width $w_{Si} = 500$ nm, height $h_{Si} = 220$ nm) that lie on a buried oxide (BOX) $SiO_2$ layer with a height of $h_{BOX} = 3\,\mu m$. Each SiP waveguide is connected to a grating coupler (GC) on one side and to a freeform waveguide on the other side. The connections between SiP waveguides and freeform waveguides are made through linear inverse tapers with a length of $l_{taper} = 60\,\mu m$ on both connected waveguides. The chosen length ensures an adiabatic transition between the different mode fields in the two waveguides and thus improves the coupling, see Figs. 3(c) and 3(d). On the SiP waveguide side, the linear taper converts the initial $w_{Si} \times h_{Si} = 500\,nm \times 220\,nm$ cross section to a $w_{Si,taper} \times h_{Si} = 130\,nm \times 220\,nm$ cross section at the taper tip, see Figs. 3(c) and 3(d). On the freeform waveguide side, the initial $w_{WG} \times h_{WG} = 2\,\mu m \times 1.8\,\mu m$ cross section is linearly tapered to a $w_{WG,taper} \times h_{WG,taper} = 0.76\,\mu m \times 1.8\,\mu m$ cross section. All freeform waveguide trajectories are plane curves with an apex height $h$ swept between $h_{min} = 6.2\,\mu m$ and $h_{max} = 16.2\,\mu m$ with a step of about 670 nm. For these experiments, we designed the beginning and the ending of each freeform waveguide trajectory (length of 22.5 μm on each side) as a circular arc, bending upwards with a radius of $R_0 = 55\,\mu m$, while the central freeform waveguide trajectory part (55 μm length) is variable, generated by a parameterized B-spline, see Figs. 3(e) and 3(g). For small values of $h$, there are two sharp bends with radius $R_1$ at the two connections between the central freeform waveguide section and the two parts with the constant bend radius, see Fig. 3(g). This bend radius $R_1$ becomes larger with increasing apex height $h$, such that the loss contribution of these bends decreases. For large values of $h$, there is a sharp bend with a radius $R_2$ at the trajectory apex, which again increases the overall loss. It is therefore expected that the transmission increases with $h$ near $h_{min}$ and decreases with $h$ near $h_{max}$, with a maximum between the two extreme values of $h$.

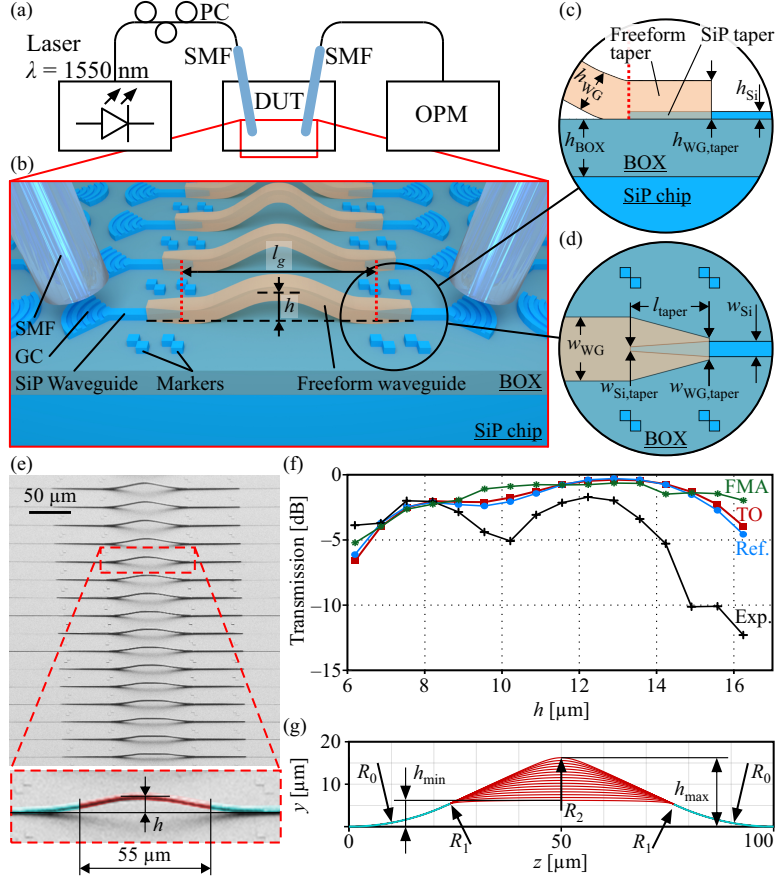

Fig. 3. Benchmarking of the TO approach with respect to different simulation techniques and to measurements. **(a)** Experimental setup for measuring the losses of freeform waveguides with different trajectories: Continuous wave (CW) light emitted by a laser source at a wavelength of $\lambda = 1550\,\text{nm}$ is launched to the device under test (DUT) consisting of 3D freeform waveguides that are connected to on-chip access waveguides. The light is coupled to the chip by single-mode fibers (SMF) and grating couplers (GC). The polarization of the incoming light is adjusted by a polarization controller (PC), and the power of the transmitted signal is measured by an optical power meter (OPM). **(b)** Artist's impression of the test structures on silicon photonic (SiP) chip: Light from an SMF enters a freeform waveguide via a GC and a SiP waveguide. The 3D freeform waveguides feature different apex heights $h$, which are measured between the trajectory in the center of the waveguide and the top surface of the buried oxide (BOX) layer. **(c)** Side view and **(d)** top view of the tapered transition between a freeform waveguide and a SiP waveguide. **(e)** Scanning electron microscope (SEM) image of a series of freeform waveguides fabricated on a SiP chip. The apex height $h_{\text{min}} \leq h \leq h_{\text{max}}$ of the trajectory is swept between $h_{\text{min}} = 6.2\,\mu\text{m}$ and $h_{\text{max}} = 16.2\,\mu\text{m}$. Bottom inset: Close-up of a freeform waveguide having an apex height of $h \approx 13\,\mu\text{m}$. The central part (red) is 55 μm long. The coupling sections and the adjacent sections (cyan) are kept the same for all freeform waveguides. **(f)** Simulated and measured transmission of the freeform waveguides for different apex heights $h$. Methods: TO — Transformation optics, Ref. — Reference simulation in original space, FMA — Fundamental-mode approximation, Exp. — Experimental. **(g)** Freeform waveguide trajectories with indicated radii of curvature $R_0$, $R_1$, and $R_2$.

To measure the transmission loss of a 3D printed freeform waveguide, continuous-wave (CW) light from a laser source at a vacuum wavelength of $\lambda = 1550\,nm$ is launched to the SiP input waveguide through a standard single-mode fiber (SMF) and a GC, Figs. 3(a) and 3(b). The light propagates through the 3D-printed freeform waveguide, and is finally coupled out through another SiP waveguide and probed by another SMF through the corresponding GC. The transmitted optical power is measured by an optical power meter (OPM). In order to exclude the coupling losses between the chip and the fibers, we measure the transmission loss of a reference structure (not shown) comprising two GC connected by a short SiP waveguide, and we refer all other transmission measurements to this value. The corresponding result is plotted on a logarithmic scale in Fig. 3(f) (marked Exp., black). To the best of our knowledge, these measurements represent the first experimental study of trajectory-dependent losses of 3D-printed freeform waveguides. The small variations in transmission for adjacent heights demonstrate the precision of the 3D printing system and the reproducibility of the overall fabrication process. Note that the cross section of 3D-printed freeform waveguide cores (2 μm × 1.8 μm) in combination with the index difference between the core ($n_{core} \approx 1.53$) and the cladding ($n_{cladding} \approx 1.36$) permits propagation not only of the fundamental mode, but also of some higher-order modes. The local transmission minimum at an apex height of $h \approx 10\,\mu m$, see Fig. 3(f), stems from the excitation of higher-order modes in the 3D freeform waveguide and from the resulting multimode interference at the transition to the single-mode SiP waveguide, see also Fig. 4.

We then simulate the freeform waveguide transmission losses with the FIT time domain solver of CST MWS using the TO approach (in the virtual $(u, v, s)$-space as described in Section 3), and with the conventional approach (in the original $(x, y, z)$-space) as a reference. We set up the ranges of $u$ and $v$ as $u \in [-2.5\,\mu m, 2.5\,\mu m]$ and $v \in [-4\,\mu m, 4\,\mu m]$ in the TO-based simulations, while in the reference simulations these ranges correspond to $x \in [-2.5\,\mu m, 2.5\,\mu m]$ and $y \in [-4\,\mu m, h + 4\,\mu m]$. Since we are predominantly interested in the impact of the waveguide trajectory on the transmission behavior, these simulations do not take into account the coupling of the SiP waveguide to the 3D-printed freeform waveguide as detailed in Figs. 3(c) and 3(d). Instead, we assume the freeform waveguide to be embedded into a homogeneous cladding material, and we cut it at the end of linear taper of the SiP waveguide, as indicated by a dotted red line in Fig. 3(c). The tapered structure to the right of this line is then replaced by a straight waveguide section of length $l_s = 1\,\mu m$, in which we define the ports for the CST MWS simulation. Since a straight waveguide in $(x, y, z)$-space maps to an identical straight waveguide in $(u, v, s)$-space without any change of material properties, the modes of the straight waveguides in both spaces are identical. The length of the computational domain in the virtual $(u, v, s)$-space is thus $s_{tot} + 2l_s$, where $s_{tot}$ denotes the total length of the freeform waveguide trajectory. For the simulation in the original $(x, y, z)$-space, the length of the computational domain is only $(l_g + 2l_s)$. All simulations were done with the same settings and on the same simulation machine. The discretization of the freeform waveguide into bricks for performing the TO simulations was done as explained in Section 3, with 28 bricks per slice as illustrated in Fig. 2. The total number of bricks for different freeform waveguides ranges from 1288 to 3080. TO-simulated transmission values in Fig. 3(f) are displayed in red, the results of the reference simulations by the conventional approach are displayed in blue. The curves obtained by the TO and the reference simulations are nearly identical, with minimal differences that can be explained by the discretization of the freeform waveguide model into finite bricks with constant material properties. Both curves, similarly to the experimentally obtained curve, exhibit a local minimum near $h = 10\,\mu m$, and agree qualitatively very well with the measurement. The deviations between measurements and simulations can be explained by the fact that the latter do not account for the coupling of the SiP waveguide to the 3D-printed freeform waveguide as detailed in Figs. 3(c) and 3(d). In addition, the higher measured transmission for $h \approx 6\,\mu m$ can be explained by a possible shrinking of the freeform waveguides during the development process, which smoothens the two sharp bends designed to have a curvature radius $R_1 \approx 4.3\,\mu m$, at the connections of the center freeform waveguide section to the initial and the final sections having a constant bend radius $R_0 = 55\,\mu m$, see Fig 3(g).

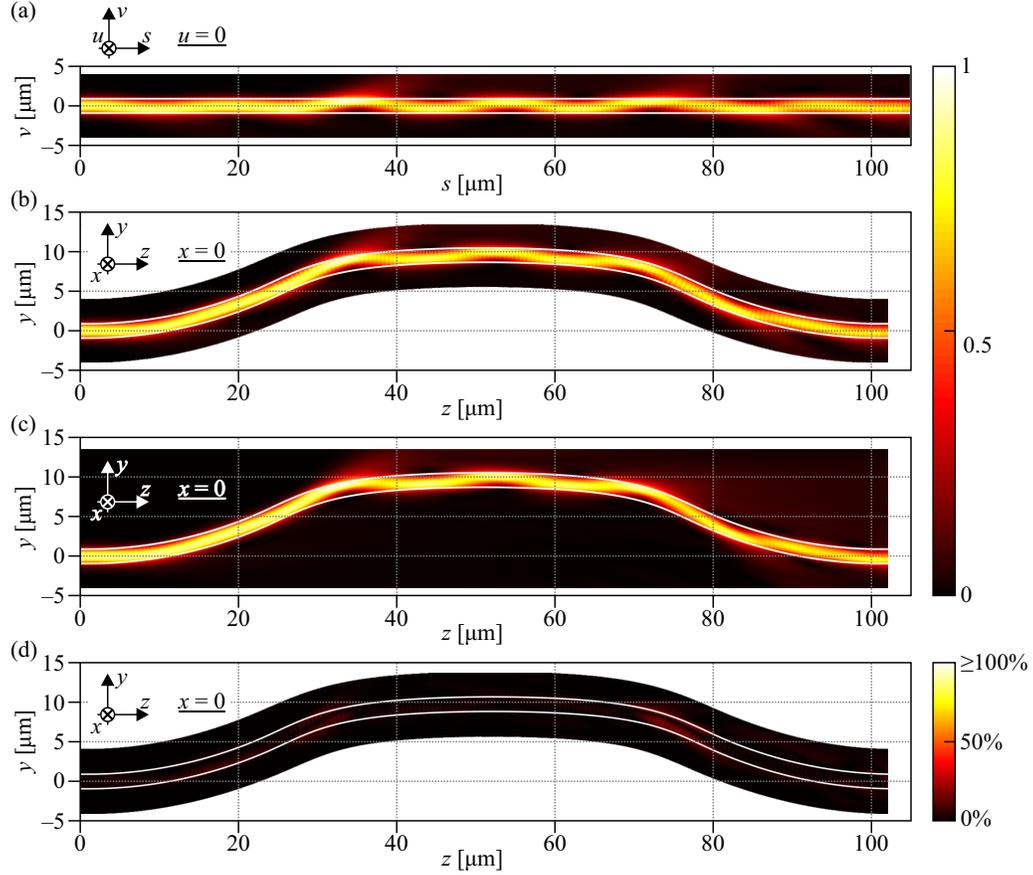

Fig. 4. Simulated normalized magnitude $|\underline{\mathbf{E}}|$ of the complex electric field vectors $\underline{\mathbf{E}}$ for the freeform waveguide with apex height $h = 9.54\,\mu\text{m}$. The complex electric field $\underline{\mathbf{E}}$ is obtained from the time-domain simulation results through a Fourier transform at the target frequency, corresponding to a vacuum wavelength of $\lambda = 1550\,\text{nm}$. We compare results in the virtual $(u, v, s)$-space and in the original $(x, y, z)$-space for a TE polarized light, having a dominant electric field oriented along $x$ in $(x, y, z)$-space and along $u$ in $(u, v, s)$-space. The freeform waveguide has a plane trajectory in the $(y, z)$-plane $(x = 0)$. The white contour lines in (a)–(d) are added for a better visualization of the freeform waveguide core. **(a)** Field distribution in $(u, v, s)$-space in plane $u = 0$, which corresponds to plane $x = 0$ in $(x, y, z)$-space. The freeform waveguide in $(u, v, s)$-space is straight, which allows to reduce the computational domain to a minimum-size rectangular volume. **(b)** Field distribution in $(x, y, z)$-space in the plane $x = 0$ obtained by applying the inverse space transformation $(x, y, z)^{\text{T}} = \mathbf{f}^{-1}(u, v, s)$ to the distribution shown in (a). **(c)** Field distribution in $(x, y, z)$-space in plane $x = 0$, as obtained from a reference simulation in $(x, y, z)$-space. The rectangular computational volume in $(x, y, z)$-space is not optimal since it comprises much space far away from the freeform waveguide, where the field is close to zero. The field distribution shows a good match to the distribution from (b). **(d)** Relative deviation between the field distributions obtained from the TO simulation in (b), and the reference simulation in (c), calculated by normalizing the magnitude of the difference of the respective electric field magnitudes to the maximum of the field magnitude found for the reference simulation. Referring to Fig. 3(f) and the field in (a)–(c), we see that multimode interference effects could explain the local minimum in transmission for apex heights $h$ around $10\,\mu\text{m}$.

We also simulate the transmission losses using the FMA method [51], which subdivides the waveguide into sections of constant curvature and calculates the transmission losses based on the propagation of the fundamental modes in these waveguide segments — the results are displayed in green in Fig. 3(f). The transmission obtained by the FMA method, does not show the local minimum near $h = 10\,\mu m$, which we attribute to the fact that the FMA method disregards possible excitation of higher-order modes, such that it cannot take into account any multi-mode interference effects. Still, the optimum configurations with the least losses around a height of $13\,\mu m$ are reasonably well reproduced. Within certain limits, the FMA can therefore be used for a real-time trajectory optimization, which is important for, e.g., photonic wire bonds [12–14, 16] or waveguide overpass fabrication [15] in an industrial setting.

For comparing the electric field distribution obtained by the TO and by the conventional approach, we plot the magnitude of the complex electric field vectors in the plane $u = 0$ of the virtual $(u, v, s)$-space, Fig. 4(a), and in the corresponding plane $x = 0$ of the original $(x, y, z)$-space after back-transformation, Fig. 4(b). The apex height of the depicted waveguide amounts to $h = 9.54\,\mu m$, which corresponds to the local minimum of the simulated transmission, caused by higher-order mode excitation. Figure 4(c) depicts the results of a CST MWS reference simulation in $(x, y, z)$-space, and Figure 4(d) displays the magnitude of the relative deviation between the field distributions obtained from the TO simulation, Fig. 4(b), and the reference simulation, Fig. 4(c), normalized to the maximum of the field magnitude found for the reference simulation. It can be seen that the two field distributions match well. The biggest differences occur in the regions with a small bend radius — these differences are attributed to the discretization of the waveguide into bricks in the transformed space, see Section 3.

## 5. Computational complexity

To quantify the advantages of the TO approach, we compare the associated computational effort to that of the conventional approach. The results are displayed in Fig 5. Figure 5(a) shows the ratios of the volumes of the computational domains for different apex heights $h$ of the freeform waveguide, see Fig. 3(b), as well as ratio of the numbers of grid cells. Both ratios indicate a reduction of the computational effort by more than a factor of two when using the TO technique. This ratio increases nearly linearly with the apex height $h$, because the computational volume of the conventional approach increases more strongly with $h$ than the one of the TO approach. Note that, for the TO-based simulations, the volume depends predominantly on the total arc length $s_{tot}$ of the freeform waveguide, which increases only slightly with $h$ as long as $h \ll s_{tot}$. The ratio of the number of grid cells for the reference simulation and the TO-simulation follows about the same proportionality as the ratio of the volumes of the computational domains. Note that the ratio of the grid cell numbers is even slightly larger than the ratio of the computational-domain volumes. We attribute this finding to the fact that, in virtual space the side walls of the straight waveguides are perfectly aligned to the rectilinear grid. Local mesh refinement as needed to accurately represent the curved waveguide surfaces in real space is hence unnecessary for the TO simulation.

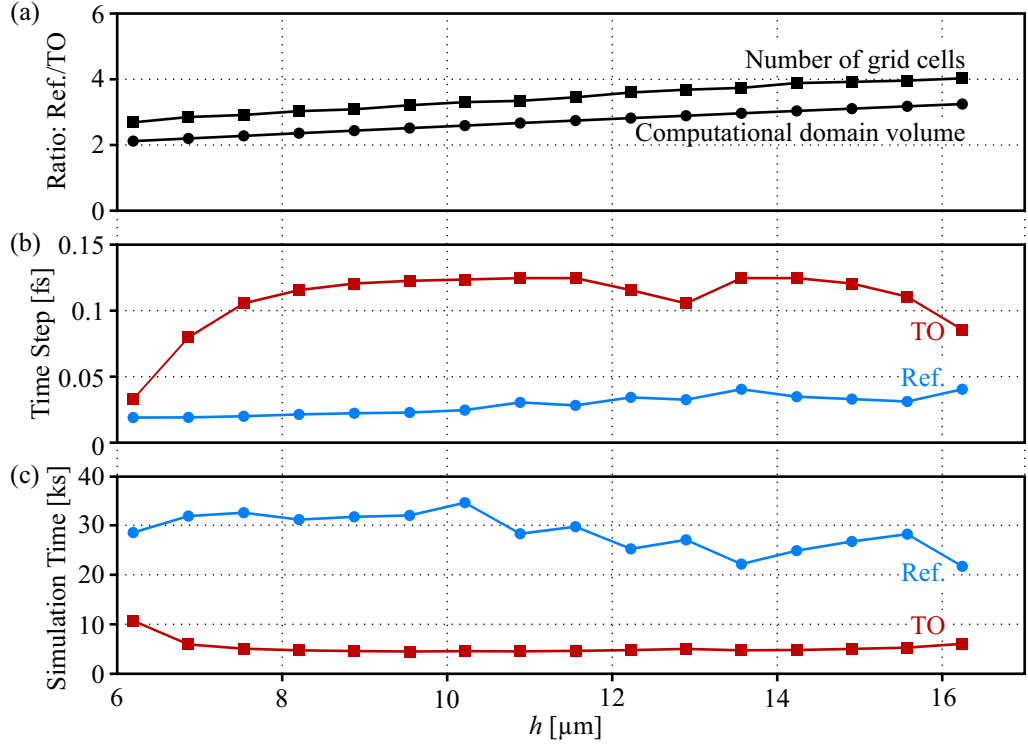

Fig. 5. Comparison of the computational effort for the TO technique and for conventional reference simulations of freeform waveguides with different apex heights $h$ of the trajectories, see Fig. 3(b). We compare the computational volume and the number of grid cells, the time steps, and the overall simulation time of the simulations in the transformed $(u, v, s)$-space (TO) to that of the direct reference simulations in the original $(x, y, z)$-space (Ref.). **(a)** Ratio of the computational domain volumes and of the corresponding numbers of grid cells for the reference simulations (Ref.) and the associated TO simulations (TO). Both curves follow a (nearly) straight line as a function of $h$. **(b)** Comparison of time steps, determined by the smallest grid cell size and by the material properties. In the conventional reference simulations, smaller grid cell sizes $\Delta x$, $\Delta y$, and $\Delta z$, are needed to correctly represent the curved surfaces of the freeform waveguides, thus leading to smaller time steps according to Eq. (7). The large variations of the time steps in the TO-based simulations originate from different material properties. Specifically, in case of freeform waveguides with sharp bends, the tensors of material properties in the transformed $(u, v, s)$-space may have elements with values close to zero, see Appendix C, which leads to large phase velocity and thus small time steps in the TO-based model. **(c)** Comparison of total simulation times. Both the different numbers of grid cells and different time steps contribute to different simulation times. Overall, for the structures simulated here, the TO approach is 3–6 times more efficient than the conventional simulations, with significant potential for further improvement. Note that the trajectories here are plane curves in the $(y, z)$-plane that start and end at the same height $y$. The advantages of the TO approach related to the volume reduction of the computational domain and thus to the total simulation time reduction, would be even more pronounced for waveguides with non-plane trajectories as, e.g., shown in Fig. 1.

Another parameter that influences the total simulation time is the *time step* used in the FIT simulation, for which the Courant-Friedrichs-Lewy stability condition for solving partial differential equations [26, 53] dictates an upper limit,

$$\Delta t_{xyz} \leq \left( c_{xyz} \sqrt{\frac{1}{\Delta x^2} + \frac{1}{\Delta y^2} + \frac{1}{\Delta z^2}} \right)^{-1}, \quad \Delta t_{uvs} \leq \left( c_{uvs} \sqrt{\frac{1}{\Delta u^2} + \frac{1}{\Delta v^2} + \frac{1}{\Delta s^2}} \right)^{-1}. \quad (7)$$

In these relations, $\Delta t_{xyz}$ is the maximal time step and $c_{xyz}$ the maximum phase velocity in any direction in $(x, y, z)$-space, which is discretized by spatial step sizes $\Delta x$, $\Delta y$, and $\Delta z$. For the TO approach, $\Delta t_{uvs}$, $c_{uvs}$ and $\Delta u$, $\Delta v$, $\Delta s$ are the equivalent quantities in $(u, v, s)$-space. Since Eq. (7) must hold for all grid cells in the computational domain, the maximum time step is eventually dictated by the smallest grid cell. In this context, the TO technique offers the additional advantage that the geometrically straight waveguide in the transformed $(u, v, s)$-space can be well represented by a rather simple rectilinear grid. In contrast to that, correct representation of the surfaces of the freeform waveguide in the original $(x, y, z)$-space may require local refinements of the grid cell sizes $\Delta x$, $\Delta y$, and $\Delta z$, and thus leads to smaller time steps according to Eq. (7). For most cases of practical interest, this advantage of the TO approach surpasses the advantage of the computational volume reduction by the transformed material tensors, see Fig. 5(b). Note, however, that the transformed material tensors $\varepsilon'$ and $\mu'$ in $(u, v, s)$-space can have elements with values close to zero, which strongly increases the associated maximum phase velocity $c_{uvs}$ and thus reduces the maximum permitted time step $\Delta t_{uvs}$ in the TO simulation. This effect occurs, e.g., towards the inner sides of strong waveguide bends, where the boundary of the computational domain may be close to a local center of curvature, see Appendix C for details. We can observe this effect indirectly from Fig. 5(b), where the shortest time step was required for the freeform waveguide with the smallest height and thus the smallest radius of curvature $R_1 \approx 4.3\,\mu\mathrm{m}$ at the transitions between the fixed and the variable part of the trajectory, see Fig. 3(g). In all simulations, the range of $v$-coordinates was $v \in [-4\,\mu\mathrm{m}, 4\,\mu\mathrm{m}]$, thus the boundary of the computational domain in this simulation was just $0.3\,\mu\mathrm{m}$ away from the center of curvature associated with the most strongly curved waveguide section. This effect could be mitigated by limiting the values of the elements of the tensors to a chosen lower bound, which will not have significant impact on the simulation results — the field is always dragged to the outer side of the bend such that the magnitude at the inner side is small. Finally, we compare the total simulation times for the TO approach and the reference simulation, see Fig. 5(c). Both results are influenced by the total number of grid cells and the time step. In all cases, the TO-based simulations were completed significantly faster than the conventional reference simulations. It should be mentioned that the additional computational overhead of the TO-based simulations, given by the time necessary to discretize the freeform waveguides into bricks, to calculate the material properties, to generate the required CST MWS scripts, and to perform the spatial back transform after the simulation, was around 10 s, while the time to execute the CST MWS scripts (to generate the the 3D models in CST MWS interface) varied from approximately 45 s to 120 s, depending on the number of bricks, which varied from 1288 to 3080. This is negligible compared to the total simulation time of several thousand seconds and was therefore not included in the results shown in Fig. 5(c). Overall, for the structures simulated here, the TO approach is 3–6 times faster than the conventional simulation, with significant potential for further improvement.

The waveguide trajectories in the presented examples are still rather simple, consisting of plane curves in the $(y, z)$-plane that start and end at the same height $y$. For waveguides with non-plane trajectories, see, e.g., Fig. 1, the reduction of the computational volume through the TO approach will be even more pronounced, but the numerical treatment also becomes more complicated. Specifically, for waveguides made from isotropic materials and having plane trajectories, the tensors $\varepsilon'$ and $\mu'$ maintain their diagonal shape in $(u, v, s)$-space, see Appendix B, while waveguides with non-plane trajectories require the consideration of off-diagonal elements in the transformed material tensors $\underline{\varepsilon}'$ and $\underline{\mu}'$. For commonly used leap-frog schemes, the FDTD update equations for the electric and the magnetic field in generally anisotropic media can be written as

$$\begin{aligned}\underline{H}'(t + \Delta t/2) &= \underline{H}'(t - \Delta t/2) - \Delta t \left(\underline{\mu}'(u, v, s)\right)^{-1} \cdot \left(\nabla \times \underline{E}'(t)\right), \\ \underline{E}'(t + \Delta t) &= \underline{E}'(t) + \Delta t \left(\underline{\varepsilon}'(u, v, s)\right)^{-1} \cdot \left(\nabla \times \underline{H}'(t + \Delta t/2)\right).\end{aligned} \quad (8)$$

In this relation, $t - \Delta t/2$, $t$, $t + \Delta t/2$ and $t + \Delta t$, denote the staggered points in time at which the electric and the magnetic fields are calculated by the leapfrog scheme, whereas $\Delta t$ denotes the corresponding time step. For isotropic materials or media with diagonal $\underline{\varepsilon}'$- and $\underline{\mu}'$-tensors in $(u, v, s)$-space, the evaluation of each of these update equations simply involves multiplication of a scalar or a (3, 1) vector by the (3, 1) vector resulting from the $\nabla \times \underline{E}'$- and the $\nabla \times \underline{H}'$-operation. Taking into account off-diagonal elements would require a multiplication of a (3, 3) matrix by the (3, 1) vector and thus increase the computational effort for the evaluation of each update equation by a factor of less than 2, see Appendix D for a more detailed explanation of the underlying model. In addition, FDTD modeling of materials with non-diagonal $\underline{\varepsilon}'$ and $\underline{\mu}'$-tensors is complicated by the coupling of non-parallel components of $\underline{H}'$ and $\nabla \times \underline{E}'$ and of $\underline{E}'$ and $\nabla \times \underline{H}'$, which are not collocated on the standard Yee grid [54]. Numerical modeling of such materials thus requires spatial interpolation [54, 55], or alternative approaches based on Lebedev grids [56, 57]. Still, taking into account that, for non-plane trajectories, the reduction of simulation time should even exceed the factor of 3–6 that we found for the case of a plane trajectory, see Fig. 5(c), we expect that the overall computational effort should still be greatly reduced by using the proposed TO approach.

Note that the simulations shown in the previous sections refer to waveguides with rectangular cross sections that are invariant along the waveguide trajectory, which can be accurately represented by a rather simple rectilinear grid. In case of arbitrary cross sections that vary along the waveguide trajectory, this advantage of the TO approach might be maintained by adapting the transformation to not only map the curved trajectory into a straight one, but to also map an arbitrary waveguide cross section into a rectangular one [58] that is invariant along the propagation direction.

## 6.  Summary

We introduced a transformation-optics (TO) approach for simulating freeform optical waveguides, which is applicable to commercially available time-domain Maxwell's equations solvers. The method reduces the computational volume by transforming a curved freeform waveguide in the original 3D space into a straight structure with a modified permittivity and permeability profile. Furthermore, the method allows for a better alignment of grid lines and surfaces of rectangular freeform waveguides, thus avoiding local mesh cell refinement. We verify the viability of our technique by simulating freeform waveguides with plane trajectories, thereby demonstrating a significant reduction in computational effort compared to the simulations in the original 3D space while maintaining the same level of accuracy. For experimental benchmarking, we realized a series of freeform waveguides and measured their transmission losses, finding good qualitative agreement. To the best of our knowledge, these measurements represent the first experimental study of trajectory-dependent losses of 3D-printed freeform waveguides. We believe that our TO-based simulation approach has the potential to greatly facilitate design and prototyping of optical devices based on 3D freeform waveguides.

## Appendix A: On the bijectivity of the coordinate transformation function

A necessary condition for the bijectivity of the coordinate transformation function $(u, v, s)^\mathrm{T} = \mathbf{f}(x, y, z)$ described by its inverse function $(x, y, z)^\mathrm{T} = \mathbf{f}^{-1}(u, v, s)$ in Eq. (3) is that the computational domain does not contain local center of curvature of the freeform waveguide trajectory. This can be achieved by appropriately choosing the ranges of $u$- and $v$-coordinates. As a simple illustrative example we may think of a freeform waveguide with a plane trajectory in the $(y, z)$-plane consisting of two straight sections that are connected by a 90° bend with a constant bend radius $R$. The local center of curvature is point $C$ in the $(y, z)$-plane, see Fig. 6(a). As explained in Section 3, the RMF in all points of the trajectory is chosen such that vector $\mathbf{U}$, pointing out of the drawing plane, is parallel to the $x$-axis and perpendicular to the $(y, z)$-plane. As a consequence, vector $\mathbf{V}$ is parallel to the $(y, z)$-plane. Since the vector $\mathbf{V}$ is parallel to the $v$-axis in $(u, v, s)$-space, the range $v \in [-d, d]$ determines whether the point $C$ is inside or outside the computational domain.

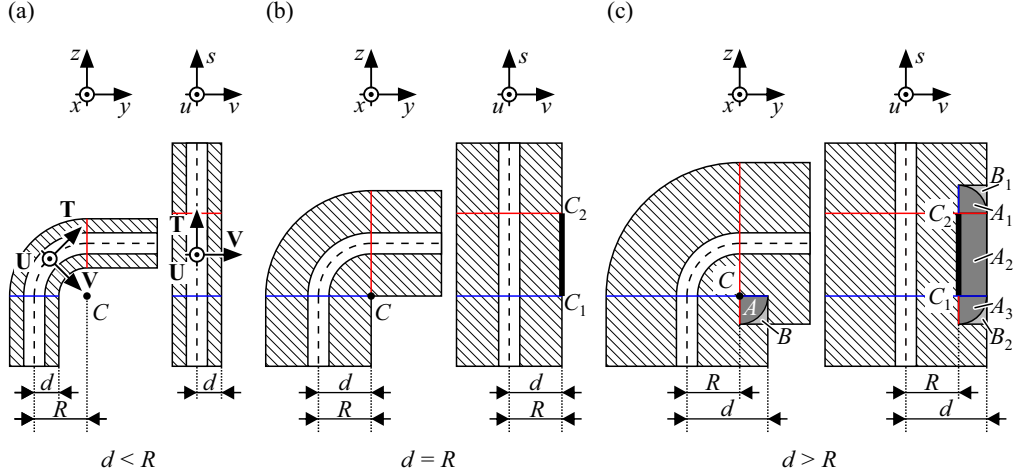

Fig. 6. Waveguide bends for illustrating a necessary condition for the bijectivity of the coordinate transformation function $(u, v, s)^T = \mathbf{f}(x, y, z)$, which is defined by its inverse function $(x, y, z)^T = \mathbf{f}^{-1}(u, v, s)$ in Eq. (3). The different computational domains are limited by the outer boundaries of the hatched areas. We consider a plane freeform waveguide trajectory (dashed lines) in the $(y, z)$-plane, consisting of two straight waveguide segments connected by a 90° circular bend with a bend radius $R$. The $(y, z)$ plane in $(x, y, z)$-space (left drawings) corresponds to the $(u, v)$-plane in $(u, v, s)$-space (right drawings). The local center of curvature corresponds to the center $C$ of the 90° bend. We consider a symmetric range of $v$-coordinates, $v \in [-d, d]$. The RMF is shown only in case (a). **(a)** If $d < R$, the point $C$ is outside the computational domain, ensuring bijective mapping between the two spaces. **(b)** If $d = R$, the point $C$ is on the border of the computational domain. In this critical case, the space transformation is not anymore a bijection, since point $C$ is mapped onto a line segment $C_1 C_2$ in $(u, v, s)$-space. **(c)** If $d > R$, the subdomains $A$ and $B$ in $(x, y, z)$-space are mapped to multiple sub-domains $A_1, A_2, A_3$ and $B_1, B_2$, respectively, in $(u, v, s)$-space, and the one-to-one correspondence between the two spaces is further violated.

Without going into mathematical details, we give a qualitative analysis of three cases: $d < R$, $d = R$, and $d > R$, Figs. 6(a)–6(c) here — a more detailed mathematical analysis of the spatial transformation of 90° bends can be found in Appendix C. Dashed lines represent waveguide trajectories, the white part in the middle represents the waveguide core, and the hatched parts to both sides represent the portion of cladding that is included in the computational domain. For all three cases we provide two drawings: One for the original freeform waveguide in the $(y, z)$-plane of $(x, y, z)$-space, and one for the corresponding straight waveguide in the $(v, s)$-plane of $(u, v, s)$-space. If $d < R$, the computational domain does not include the center point $C$ of the bend, and the mapping between the two spaces is a bijection, see Fig. 6(a). In case $d = R$, point C is on the border of the computational domain and is mapped to a line segment $C_1C_2$ in $(u, v, s)$-space, the length of which is equal to the arc-length of the 90° bend of the trajectory, $\overline{C_1C_2} = R\pi/2$, see Fig. 6(b). This is the critical case when the spatial transformation function is not a bijection anymore. In case $d > R$, the one-to-one correspondence between the two spaces is further violated. Not only the point $C$ is again mapped to the line segment $C_1C_2$, but also areas marked with $A$ and $B$ in the $(y, z)$-plane are mapped into multiple areas in the $(v, s)$-plane — see Fig. 6(c). In particular, area $A$ is mapped to areas $A_1$, $A_2$, and $A_3$, while area $B$ is mapped to areas $B_1$ and $B_2$. Based on this simple example, it is clear that a necessary condition for ensuring bijectivity is that $C$ in Fig. 6 is outside of the computational domain. On the other hand, for a physically correct representation of the waveguide, the essential part of the evanescent fields at the outside of the bend must be inside the computational domain. This might be accomplished by choosing a computational domain that comprises an asymmetric range of $v$-coordinates instead of the symmetric range $[-d, d]$ used in our example.

This simple example is also very illustrative for the general case of freeform waveguides. Bends need not necessarily be 90° bends, and the bend radius can continuously change along the trajectory. In case of non-plane true 3D trajectories, the local center of the bend lies in the local osculating plane, and the same reasoning provided above can be applied. As a matter of fact, the osculating plane for our example is the $(y, z)$-plane — the only difference to 3D trajectories would be that the local vectors of the RMF are not necessarily perpendicular and parallel to the local osculating plane for each point on the trajectory.

## Appendix B: Tensors of $\varepsilon'$ and $\mu'$ in $(u, v, s)$-space for freeform waveguides with plane trajectories

In case of freeform waveguides with plane trajectories and isotropic material properties $\varepsilon$ and $\mu$ in the original $(x, y, z)$-space, the material properties $\varepsilon'$ and $\mu'$ in the virtual $(u, v, s)$-space are diagonal tensors. This can be shown by assuming that, without loss of generality, the plane trajectory lies in $(y, z)$-plane, as explained in Section 3 and Appendix A, such that the vector $\mathbf{U}$ of the RMF in each trajectory point is oriented parallel to the $x$-axis, while the remaining two vectors $\mathbf{V}$ and $\mathbf{T}$ of the RMF lie in the $(y, z)$-plane, with the vector $\mathbf{T}$ forming an angle $\theta$ with the positive $z$-axis, see Fig. (7). Since the vectors $\mathbf{V}$ and $\mathbf{T}$ are parallel to the $v$- and $s$-axes in the $(u, v, s)$-space, respectively, the axes of the 2D $(v, s)$-coordinate system are rotated by the same angle $\theta$ with respect to the axes of the 2D $(y, z)$-coordinate system. This allows us to simplify the calculation of the Jacobian matrix given by Eq. (2). Since the vector $\mathbf{U}$ and the $x$-axis are parallel to each other, it follows that $\partial u/\partial x = 1$. This also implies that the partial derivatives of $u$ with respect to $y$ and $z$ must be zero, $\partial u/\partial y = 0$, and $\partial u/\partial z = 0$. Furthermore, the vectors $\mathbf{V}$ and $\mathbf{T}$ are perpendicular to the $x$-axis, which implies that the derivatives of $v$ and $s$ with respect to $x$ must be zero, too: $\partial v/\partial x = 0$ and $\partial s/\partial x = 0$. The Jacobian matrix of the coordinate transformation function thus reads

$$\mathbf{J} = \begin{bmatrix} 1 & 0 & 0 \\ 0 & \frac{\partial v}{\partial y} & \frac{\partial v}{\partial z} \\ 0 & \frac{\partial s}{\partial y} & \frac{\partial s}{\partial z} \end{bmatrix}. \tag{9}$$

Assuming the material properties are isotropic in $(x, y, z)$-space, Eq. (1) can be simplified,

$$\begin{aligned} \varepsilon'(u, v, s) &= \varepsilon(x, y, z) \frac{\mathbf{J} \cdot \mathbf{J}^{\mathrm{T}}}{\det(\mathbf{J})}, \\ \mu'(u, v, s) &= \mu(x, y, z) \frac{\mathbf{J} \cdot \mathbf{J}^{\mathrm{T}}}{\det(\mathbf{J})}, \end{aligned} \tag{10}$$

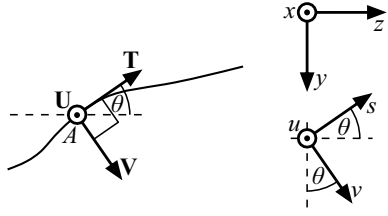

Fig. 7. Relationship between coordinates in $(x, y, z)$- and $(u, v, s)$-space in a point $A$ with coordinates $(y_0(s), z_0(s))$ on the trajectory for the case of a freeform waveguide with a plane trajectory. The trajectory lies in the $(y, z)$-plane, and the RMF is oriented such that the vector $\mathbf{U}$ and the associated $u$-axis are parallel to the $x$-axis in all points of the trajectory. The remaining two axes $\mathbf{V}$ and $\mathbf{T}$ of the RMF define a 2D frame that lies in the $(y, z)$-plane. The axes of the local $(v, s)$-coordinate system are parallel to the $(\mathbf{V}, \mathbf{T})$ frame and rotated by an angle $\theta$ with respect to the axes of the $(y, z)$-coordinate system.

where the product of the Jacobian matrix and its transposed reads

$$\mathbf{J} \cdot \mathbf{J}^\mathbf{T} = \begin{bmatrix} 1 & 0 & 0 \\ 0 & \left(\frac{\partial v}{\partial y}\right)^2 + \left(\frac{\partial v}{\partial z}\right)^2 & \left(\frac{\partial y}{\partial v}\frac{\partial y}{\partial s}\right)^{-1} + \left(\frac{\partial z}{\partial v}\frac{\partial z}{\partial s}\right)^{-1} \\ 0 & \left(\frac{\partial y}{\partial v}\frac{\partial y}{\partial s}\right)^{-1} + \left(\frac{\partial z}{\partial v}\frac{\partial z}{\partial s}\right)^{-1} & \left(\frac{\partial s}{\partial y}\right)^2 + \left(\frac{\partial s}{\partial z}\right)^2 \end{bmatrix}. \qquad (11)$$

Equation (11) is a diagonal matrix if its two off-diagonal elements are equal to zero, which reduces to

$$\frac{\partial y}{\partial v}\frac{\partial y}{\partial s} + \frac{\partial z}{\partial v}\frac{\partial z}{\partial s} = 0. \qquad (12)$$

Since $\varepsilon(x, y, z)$, $\mu(x, y, z)$, and $\det(\mathbf{J})$ are scalars, it follows from Eqs. (10) and (11) that Eq. (12) is a sufficient condition that ensures that $\varepsilon'(u, v, s)$ and $\mu'(u, v, s)$ are diagonal tensors.

The coordinate transformation from the $(v, s)$-coordinate system to the $(y, z)$-coordinate system can be extracted from the sketches in Fig. (7),

$$\begin{bmatrix} y(s, v) \\ z(s, v) \end{bmatrix} = \begin{bmatrix} y_0(s) \\ z_0(s) \end{bmatrix} + \begin{bmatrix} v \cos \theta(s) \\ v \sin \theta(s) \end{bmatrix}, \qquad (13)$$

where $(y_0(s), (z_0(s))$ are coordinates of the point $A$ on the trajectory that is defined by arc-length coordinate $s$. Equation (12) can be expressed as a dot-product of two $(2, 1)$-vectors,

$$\frac{\partial}{\partial v}\begin{bmatrix} y(s, v) \\ z(s, v) \end{bmatrix} \cdot \frac{\partial}{\partial s}\begin{bmatrix} y(s, v) \\ z(s, v) \end{bmatrix} = 0. \qquad (14)$$

Inserting Eq. (13) into Eq. (14), we obtain

$$\begin{bmatrix} \cos \theta(s) \\ \sin \theta(s) \end{bmatrix} \cdot \left( \mathbf{T}(s) + \begin{bmatrix} -v \sin \theta(s) \frac{\partial \theta(s)}{\partial s} \\ v \cos \theta(s) \frac{\partial \theta(s)}{\partial s} \end{bmatrix} \right) = \mathbf{N}(s) \cdot \mathbf{T}(s) \left( 1 + v \frac{\partial \theta(s)}{\partial s} \right) = 0, \qquad (15)$$

where $\mathbf{N}(s)$ denotes the unit normal vector in point $A$ on the trajectory, which in case of plane trajectories is parallel to vector $\mathbf{V}(s)$. Since the dot product of the normal and the tangent vector is always equal to zero, Eq. (15) is always fulfilled, and $\mu'(u, v, s)$ are thus diagonal tensors.

## Appendix C: Time-stepping

We have already shown in Appendix A that no local center of curvature of the trajectory is allowed to be within the computational domain to maintain the bijectivity of the space transformation defined by Eq. (3). In addition, if the computational domain border is too close to a local center of curvature, some elements of the tensors $\varepsilon'(u, v, s)$ and $\mu'(u, v, s)$ can assume values close to zero. The Courant-Friedrichs-Lewy stability condition, Eq. (7), then leads to small time steps of the corresponding time-domain simulation and greatly increases the overall simulation time. For illustration and similarly to Appendix A, we discuss a circular $90°$-bend with radius $R$ in the $(y, z)$-plane, see Fig. 8. Assuming, without loss of generality, that the center of the circular arc is in the origin of the $(x, y, z)$ coordinate system, the trajectory is given by

$$\begin{aligned} x_0 &= 0, \\ y_0 &= -R\cos(t), \\ z_0 &= R\sin(t), \end{aligned} \tag{16}$$

with $t \in [0, \pi/2]$. The coordinate $s = Rt$ is the arc length of the trajectory. Since the $u$-axis is parallel to the $x$-axis, and the $v$-axis is in the plane of the trajectory, we can write the one-to-one correspondence between the $(x, y, z)$- and $(u, v, s)$-spaces, see Fig. 8,

$$\begin{aligned} x(u, v, s) &= u, \\ y(u, v, s) &= -(R - v)\cos\left(\frac{s}{R}\right), \\ z(u, v, s) &= (R - v)\sin\left(\frac{s}{R}\right). \end{aligned} \tag{17}$$

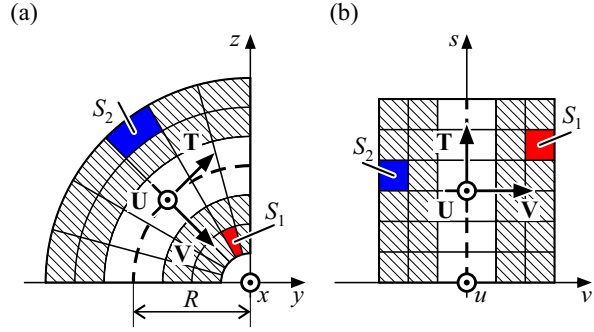

Fig. 8. Illustration of the space transformation for a $90°$-bend. **(a)** $90°$-bend waveguide in the $(y, z)$-plane of $(x, y, z)$-space. **(b)** Same waveguide in the $(v, s)$-plane of $(u, v, s)$-space. In $(u, v, s)$-space, the $90°$-bend is straightened. It is divided into squares by taking equidistant divisions along the $s$-axis (slicing) and along the $v$-axis. The squares in $(u, v, s)$-space correspond to sections bounded by two line segments and two concentric arcs in $(x, y, z)$-space. The areas of these sections differ along the radial coordinate and are smaller on the inner side and larger on the outer side of the bend. Areas close to the origin tend to zero, and mapping these sections to finite-size squares in $(u, v, s)$-space causes two elements of tensors of material properties to tend to zero, see Eq. (19). This leads to a small time step of the corresponding FDTD simulation due to the Courant-Friedrichs-Lewy stability condition, Eq. (7), and thus to long simulation times. Areas of sections far from the origin tend to infinity, and mapping infinite sections to finite size squares in $(u, v, s)$-space also causes one of the tensor elements of material properties to tend to zero. These cases are, however, not critical, since the space far away from the trajectory is commonly not of interest for modeling of 3D freeform waveguides.

The Jacobian $\mathbf{J}$ of the function $(u, v, s)^T = \mathbf{f}(x, y, z)$ can be found as the inverse of the Jacobian of the function $(x, y, z)^T = \mathbf{f}^{-1}(u, v, s)$,

$$\mathbf{J} = \begin{bmatrix} \frac{\partial x}{\partial u} & \frac{\partial x}{\partial v} & \frac{\partial x}{\partial s} \\ \frac{\partial y}{\partial u} & \frac{\partial y}{\partial v} & \frac{\partial y}{\partial s} \\ \frac{\partial z}{\partial u} & \frac{\partial z}{\partial v} & \frac{\partial z}{\partial s} \end{bmatrix}^{-1} = \begin{bmatrix} 1 & 0 & 0 \\ 0 & \cos\left(\frac{s}{R}\right) & -\sin\left(\frac{s}{R}\right) \\ 0 & \frac{R}{R-v}\sin\left(\frac{s}{R}\right) & \frac{R}{R-v}\cos\left(\frac{s}{R}\right) \end{bmatrix}. \tag{18}$$

Inserting this result into Eq. (10) leads to

$$\varepsilon'(u, v, s) = \varepsilon(x, y, z) \begin{bmatrix} \frac{R-v}{R} & 0 & 0 \\ 0 & \frac{R-v}{R} & 0 \\ 0 & 0 & \frac{R}{R-v} \end{bmatrix} = \begin{bmatrix} \varepsilon_{1,1} & 0 & 0 \\ 0 & \varepsilon_{2,2} & 0 \\ 0 & 0 & \varepsilon_{3,3} \end{bmatrix}. \tag{19}$$

A similar result can be obtained for $\boldsymbol{\mu}'$. Hence, when $v$ approaches $R$, $\varepsilon_{1,1}$ and $\varepsilon_{2,2}$ ($\mu_{1,1}$ and $\mu_{2,2}$) tend to 0, causing the maximum phase velocity to tend to infinity. According to the Courant-Friedrichs-Lewy stability condition Eq. (7), the maximal time step then tends to zero, and the total simulation time approaches infinity. On the outer side of the bend, for $v < 0$, tensor elements $\varepsilon_{3,3}$ and $\mu_{3,3}$ are more problematic, since they tend to zero for $v \to -\infty$. This is, however, not of practical relevance since the space far away from the trajectory is commonly not of interest for modeling of 3D freeform waveguides.

## Appendix D: Impact of non-plane trajectories on computational complexity

In general, freeform waveguide trajectories are non-plane, which leads to non-diagonal tensors $\boldsymbol{\mu}'$ and $\varepsilon'$ in the transformed $(u, v, s)$-space. This increases the computational complexity, as more additions and multiplications need to be done in each time step of the Maxwell's equations solver compared to the case with diagonal tensors $\boldsymbol{\mu}'$ and $\varepsilon'$. As an example, we will consider an FDTD solver, which is based on the leap-frog algorithm, where in each time step, the electric field $\underline{\mathbf{E}}'$ and the magnetic field $\underline{\mathbf{H}}'$ are updated successively. The update relation is given by Eq. (8), that we repeat here for convenience:

$$\begin{aligned} \underline{\mathbf{H}}'(t + \Delta t/2) &= \underline{\mathbf{H}}'(t - \Delta t/2) - \Delta t \left(\boldsymbol{\mu}'(u, v, s)\right)^{-1} \cdot \left(\nabla \times \underline{\mathbf{E}}'(t)\right), \\ \underline{\mathbf{E}}'(t + \Delta t) &= \underline{\mathbf{E}}'(t) + \Delta t \left(\varepsilon'(u, v, s)\right)^{-1} \cdot \left(\nabla \times \underline{\mathbf{H}}'(t + \Delta t/2)\right). \end{aligned} \tag{20}$$

For updating the vector of the electric (magnetic) field in Eq. (20), we need to calculate a curl of the magnetic (electric) field vector, multiply the $(3, 3)$ matrix $\Delta t \left(\varepsilon'\right)^{-1}$ ($\Delta t \left(\boldsymbol{\mu}'\right)^{-1}$) by the calculated curl, and add the result to the vector of the electric (magnetic) field in the previous time point. To find a curl of the $(3, 1)$ vector, we need to calculate six partial derivatives. This requires nine additions — six additions ($6A$) to calculate the finite differences between field components in spatial grid points and three more additions ($3A$) for calculating the difference between the two spatial derivatives in each component of the curl — and six multiplications ($6M$) for dividing the finite differences of the fields by the corresponding size of the spatial grid cell. In case of plane trajectories and isotropic materials, the transformed permittivity and permeability tensors $\varepsilon'$ and $\boldsymbol{\mu}'$ are diagonal, and the same applies to the $(3, 3)$ matrices $\Delta t \left(\varepsilon'\right)^{-1}$ and $\Delta t \left(\boldsymbol{\mu}'\right)^{-1}$. In this case, multiplication of these matrices by the curl vector requires three additional scalar multiplications ($3M$). In contrast to that, non-plane trajectories lead to non-diagonal tensors, for which a full vector-matrix multiplication with an overall nine multiplications and six additions is needed ($6A + 9M$). Finally, in both cases, adding the resulting vector to the field vector in the previous time point requires three more additions ($3A$). Hence, updating the electric or the magnetic field in one spatial grid point requires twelve additions and nine multiplications $12A + 9M$ in case of plane trajectories and diagonal permittivity and permeability tensors, and eighteen additions and fifteen multiplications $18A + 15M$ in case of non-diagonal tensors. The ratio of necessary arithmetic operations is therefore $\frac{18A+15M}{12A+9M} < \frac{5}{3}$. Based on this rather simple model for computational effort, we estimate that moving from plane trajectories to non-plane trajectories increases the computational effort by a factor of less than 2. Note that additional complications arise from the fact that the standard Yee grid cannot be used for non-diagonal permittivity and permeability tensors [54]. On the other hand, the reduction of simulation time for non-plane trajectories should even exceed the factor of 3–6 that we found for the case of a plane trajectory, see Fig. 5(c). We thus expect that the overall computational effort should still be greatly reduced by using the proposed TO approach

**Funding**

Deutsche Forschungsgemeinschaft (DFG), CRC 1173 "Wave Phenomena", Project C4 (258734477); DFG under Germany's Excellence Strategy via the Excellence Cluster 3D Matter Made to Order (EXC-2082/1, 390761711); Erasmus Mundus Joint Doctorate Program EUROPHOTONICS (159224-1-2009-1-FR-ERA MUNDUS-EMJD); European Research Council (ERC) Consolidator Grant "TeraSHAPE" (773248); Bundesministerium für Bildung und Forschung (BMBF) Project DiFeMiS (16ES0948); Karlsruhe School of Optics and Photonics (KSOP); Alfried Krupp von Bohlen und Halbach Foundation